# Consequences of anisotropy in electrical charge storage: application to the characterization by the mirror method of TiO$_2$ rutile.

G. Damamme[1], C. Guerret-Piécourt[2], T. Temga[2], D. Juvé[2], D. Tréheux[2]

[1] CEA /DAM Ile de France, BP 12, 91680 Bruyères-le-Chatel, France
[2] Laboratoire de Tribologie et Dynamique des Systèmes, UMR CNRS 5513, Ecole Centrale de Lyon, 36 avenue Guy de Collongue, 69134 Ecully cedex France

E-mail : christelle.guerret@ec-lyon.fr

**Abstract**:
This article is devoted first to anisotropic distributions of stored electric charges in isotropic materials, second to charge trapping and induced electrostatic potential in anisotropic dielectrics.
On the one hand, we examine the case of anisotropic trapped charge distributions in linear homogeneous isotropic (LHI) insulators, obtained after an electron irradiation in a scanning electron microscope. This injection leads to the formation of a mirror image. We first establish the characteristics of the mirror image obtained from such anisotropic distribution by linking the mirror diameter to the curvature tensor of the equipotentials thanks to the geometric optics ansatz (GOA). Second, the equipotentials induced by the presence of an anisotropic charge distribution in such isotropic dielectrics have been determined in the case of homeoidal (ellipsoidal) distributions that generalize the classical spherical distributions. Then, for these homeoidal distributions in isotropic dielectrics, the features of the mirror image have been deduced from the previous GOA estimation. Elliptic mirrors can be obtained and calculated in the limit cases of such homeoidal distributions.
On the other hand, we consider the non-trivial case of a point charge lying at the interface between the vacuum and a linear homogeneous orthotropic anisotropic (LHOA) dielectric and the determination of its corresponding potential seen from the vacuum. This problem has soon been solved in the case of transversal isotropic (TI) dielectric ($\varepsilon_x = \varepsilon_y = \varepsilon_r$, $\varepsilon_r \neq \varepsilon_z$), but we extend in this paper the classical dielectric image problem to the more general case where $\varepsilon_x \neq \varepsilon_y \neq \varepsilon_z$. The equivalent charge and the induced electrostatic potential are evaluated. For these anisotropic insulators, the equipotentials created by a point charge at the interface are found to be ellipsoids and this leads to an elliptic mirror image. The ratio between the two main axis values of the elliptic mirror is proportional to the square root of the ratio of the permittivities values in the plane of the interface. Finally these calculations are used to explain the experimental results obtained by the mirror method on TiO$_2$ sample that is known to be an anisotropic dielectric.

## 1. Introduction

Technology relevant materials exhibit very often anisotropic dielectric permittivity. Among these materials we find intrinsically anisotropic crystals such as the wide used rutile (TiO$_2$) or quartz. Some materials become also anisotropic because of their deposition or elaboration process, for example strained layers obtained after epitaxial growth [1] or other deposition process [2], or films obtained by anisotropic elaboration process such as polymer film stretching. Moreover the present interest in nanomaterials has also lead to the emergence of new anisotropic dielectrics such as filled polymers or block copolymers [3]. Some of these anisotropic materials are supposed to be good candidates in some applications where the charge storage has to be controlled, for example in thin film insulating spacecraft materials subjected to space radiations.

It is thus interesting to be able to characterize properly the ability of these anisotropic materials to trap electrical charges, and to know the electrostatic potential induced by the charge storage.

Among the various methods available to characterize this charge storage or transport ability [4], the scanning electron microscope mirror effect (SEMME) can be used to determine charge trapping under electron beam irradiation [5, 6]. The SEMME method is composed of two steps: (i) electrons of high energy (some tens of keV) are punctually injected in an insulator. The trapped electrons create an electrostatic field outside the sample. (ii) When this sample is observed at low voltage (some hundreds of eV) with a lecture scanning e-beam, the previous field can be sufficiently high to deflect the low energy electrons as a convex mirror with the light. The obtained mirror image displays a distorted view of the SEM chamber (cf. figure 1). Generally the mirror image is circular. By using the geometric optics ansatz (GOA), it is possible to link the apparent radius of the mirror image to the total trapped charge in the case of a linear homogeneous isotropic dielectric (LHI where $\varepsilon_x = \varepsilon_y = \varepsilon_z$). This GOA has been initially applied to determine the relation between the mirror plot and the curvature of the equipotentials induced by the trapped charges in the case of a point charge distribution [5]. However some elliptic mirror images have been also experimentally observed on anisotropic insulators or on samples that have been submitted to various tests such as friction test or bending test [7]. One interpretation of these elliptic mirrors could be the anisotropy of the charge distribution or the anisotropy of the dielectric.

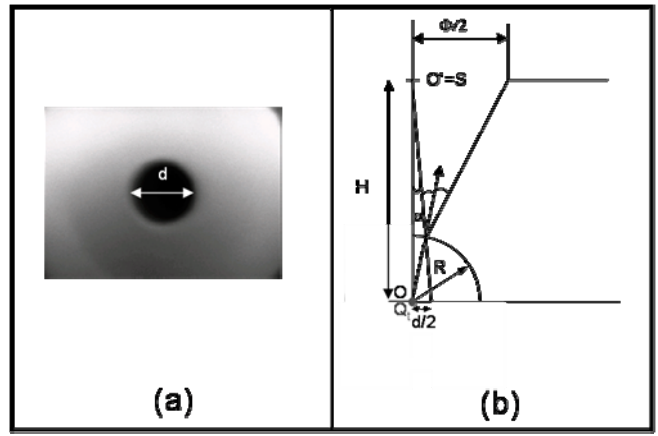

**Figure 1.** Mirror effect after an electron injection in an insulator: (a) SEM image of the injected area. The black disk (diameter d) at the center of the image corresponds to the last output diaphragm image (b) Representation of the reflection by the stored charges $Q_t$ of the low voltage accelerated electrons of the lecture e-beam: H working distance, R curvature radius of the equipotential, Φ real diameter of the last output diaphragm.

In the first part of this paper (§ 2), we use this geometric optics ansatz to determine the mirror image features, for any shape of the equipotentials and so for any trapped charge distribution.

Second (§ 3), we consider the problem of the anisotropy of the trapped charge distribution. The problem of elliptical (more precisely homeoidal) charges distributions is initially considered and solved in the case of linear homogeneous isotropic materials. In this case, the potential created by the anisotropic charge distribution is determined in the two limit cases of a charged disk perpendicular to the vacuum/dielectric interface and a charged segment in the injection plane. Then the geometrical features of the corresponding mirror images are calculated.

In a third part (§ 4), the case of anisotropic material is considered. We construct an approximate evaluation of the equivalent charge and potential seen from the vacuum when a charge is placed at the vacuum/dielectric interface for a linear homogeneous orthotropic anisotropic dielectric (LHOA, where $\varepsilon_x \neq \varepsilon_y \neq \varepsilon_z$). Finally, starting from the already resolved case of a point charge internal to a linear homogeneous isotropic-transverse dielectric (LHIT, where $\varepsilon_x = \varepsilon_y \neq \varepsilon_z$), solutions of equivalent charges and electrostatic potential are deduced for homeoidal distributions in the both cases of isotropic-transverse (LHIT) and orthotropic anisotropic dielectrics (LHOA). Comparisons of the results obtained by the analytical calculations with finite element simulations are provided.

In the last part of the article (§ 5), thanks to the previous calculations, the experimental evolution of the mirror curve obtained on $TiO_2$ samples are explained. The trapped charges are evaluated and the shapes of the trapped charges distributions are determined in the case of this anisotropic material, whose mirror images can sometimes be elliptic.

## 2. Backscattering phenomena due to trapped charges: generalization of the geometric optics ansatz (GOA)

### *2.1. Recalls on the mirror effect in LHI dielectric*

After a punctually irradiation with an electron beam of high energy (some tens of keV), electrical charges are created and/or trapped in the insulator. These charges produce an electric field in the vacuum chamber of the Scanning Electron Microscope (SEM). If the sample is then observed with an electron beam of low acceleration lecture potential $V_l$ (corresponding to an energy of some hundreds of eV), the field can backscatter the incident electrons as a convex mirror with the light. As a consequence, a mirror image appears on the SEM screen that is a distorted view of the SEM chamber and more particularly of the last output diaphragm (figure 1(a)).

Let us consider first that the trapped charge is a point charge $Q_t$ trapped in O as seen in figure 1(b). The primary electrons of the e-beam are emitted with a small angle $\alpha$ ($\alpha \cdot OO' = d/2$) with respect to the optical axis OO' from a point source S that coincides with the center O' of the last output diaphragm of the optical column of the SEM. Moreover if the minimum distance between the beam and the point charge located in O is $r_m \ll OO'$ then the beam fulfilled the Gaussian approximation. Under this approximation it has been previously demonstrated by Vallayer & al that the diameter d of the black spot of the mirror image is given by the equation (1) [5]:

$$\frac{1}{d} = \frac{4H}{\phi} \frac{2\pi\varepsilon_0(\varepsilon_r + 1)}{Q_t} V_l \qquad (1)$$

where $\phi$ is the diameter of the last output diaphragm of the SEM, H=OO' is the working distance, $\varepsilon_r$ is the dielectric permittivity of the LHI, and $V_l$ the lecture potential. The term $\frac{2Q_t}{1+\varepsilon_r}$ is the equivalent charge seen from the vacuum. One can note that this equation has been only established when the equipotentials are spherical like in the point charge approximation and for a linear homogeneous isotropic dielectric. Under these hypotheses of a LHI dielectric and of a point charge distribution, the mirror image is circular. If the stored charge distribution is no more punctual but still axisymetric compared to the e-injection direction, the mirror image stays circular, and some calculations have been provided by Attard & al to link the mirror image to the trapped charge distribution[8]

### *2.2. Geometric optics ansatz for a charge distribution of any shape*

For a charge distribution of any shape and the resulting equipotentials, the hypothesis (ansatz) will be assumed that the backscattering of an electron with a kinetic energy $E_c$ is equivalent to the optical reflection of the electron trajectory on the equipotential $V = 2E_c/e = 2V_l$ of local curvature $1/R$ (figure 2). This hypothesis is fully justified in the case of a point charge. Indeed, in this case the exact calculation is possible and leads to the same result as the GOA. This ansatz could be extended to any distribution charge system, but the relation between the apparent diameter d of the mirror image and the local curvature of the equipotential has to be determined.

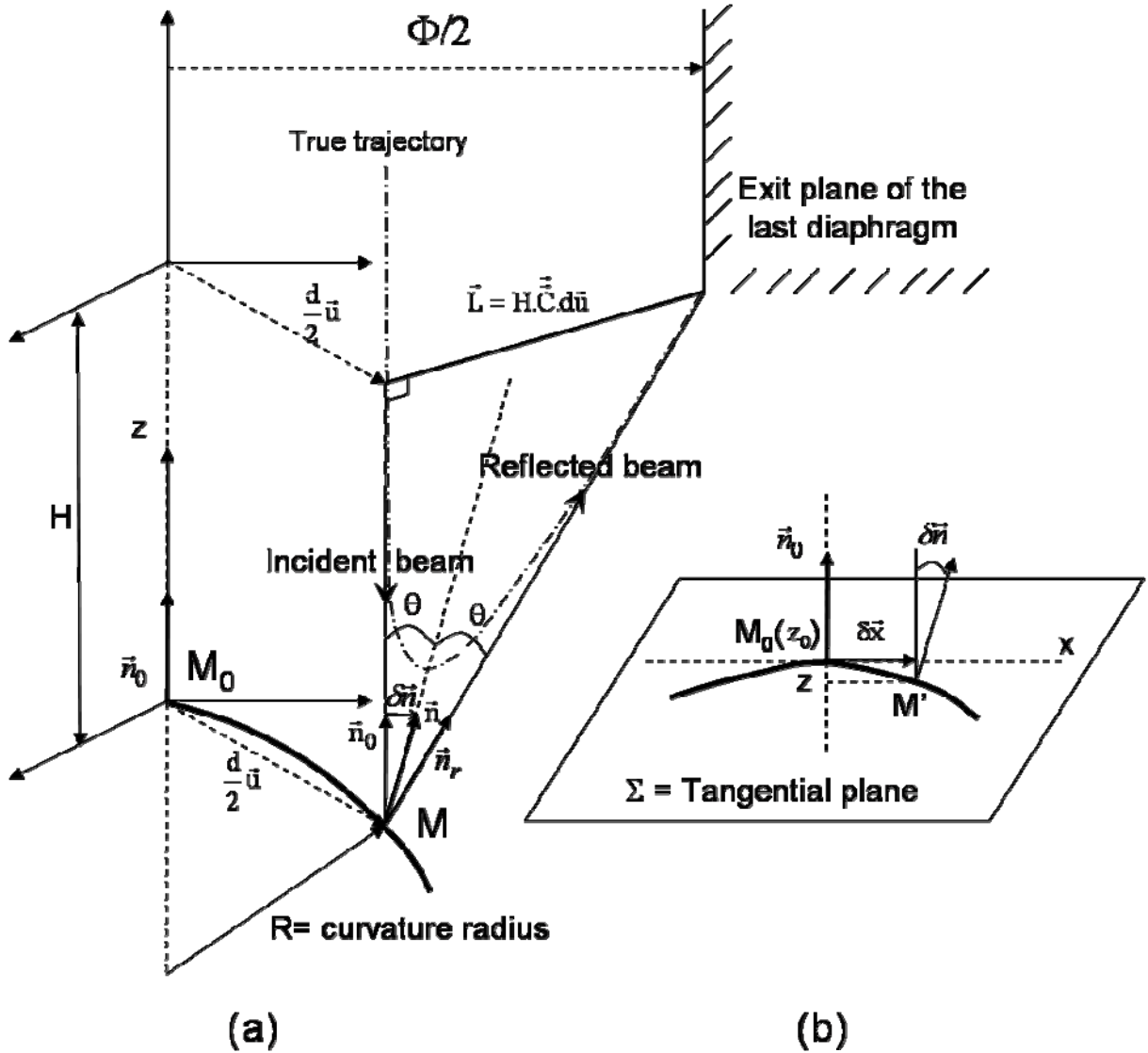

**Figure 2.** (a) Geometrical representation of the electron diffusion phenomenon in a mirror experiment, R curvature radius of the equipotential (b) Notations in the tangential plane of the equipotential.

*2.3. Mirror characteristic in the case of an equipotential of any curvature*
On figures 2(a) and 2(b), the geometric features and notations used in the following are detailed. We note $M_0(\vec{x}_0)$ the impact point of the incident electron beam on an equipotential surface of any shape characterized by its curvature tensor $\vec{\vec{C}}_\Sigma$. The tensor of curvature $\vec{\vec{C}}_\Sigma$ of the equipotential is defined thanks to his principal curvature radii $R_1$ and $R_2$ that depend on the potential value $V = 2V_1$:

$$\vec{\vec{C}}_\Sigma = \begin{pmatrix} \dfrac{1}{R_1} & 0 \\ 0 & \dfrac{1}{R_2} \end{pmatrix} \quad (2)$$

The curvature tensor gives the deviation of the surface normal vector $\vec{n}$ versus the distance $\delta\vec{x}$ in the tangential plane $\Sigma$. So, for the incident e-beam corresponding to a reflected beam that reaches the limit of the output diaphragm (figure. 2(a))

$$\vec{n} = \vec{n}_0 + \delta\vec{n} = \vec{n}_0 + \vec{\vec{C}}_\Sigma \cdot \delta\vec{x} = \vec{n}_0 + \frac{d}{2}\vec{\vec{C}} \cdot \vec{u} \quad (3)$$

where d = 2R$_{mirror}$ corresponds, as previously defined, to the apparent diameter of the black spot on the mirror image. The direction of the incident e-beam corresponds to -$\vec{n}_0$ and, with the small angle approximation, the direction of the reflected e-beam is given by

$$\vec{n}_r = \vec{n}_0 + 2\delta\vec{n} = \vec{n}_0 + d\vec{C}_\Sigma.\vec{u} \tag{4}$$

With the approximation that HC$_\Sigma \gg 1$, we can write now that the impact point of the reflected beam on the plane of the last diaphragm is equal to the radius of the last diaphragm $\Phi/2$:

$$\left\|\frac{d}{2}\vec{u} + H\vec{C}_\Sigma.d\vec{u}\right\| \approx \left\|H\vec{C}_\Sigma.d\vec{u}\right\| = \frac{\Phi}{2} \tag{5}$$

This leads to the equation:

$$\frac{1}{d} = \frac{2}{\Phi}H\sqrt{{}^t\vec{u}.C_\Sigma^2.\vec{u}} \tag{6}$$

that is equivalent to the equation:

$$^t\vec{d}.C_\Sigma^2.\vec{d} = \frac{d_x^2}{R_1^2} + \frac{d_y^2}{R_2^2} = \left(\frac{\Phi}{2H}\right)^2 \tag{7}$$

where $\vec{d}/2 = \frac{d}{2}\vec{u} = \begin{pmatrix} \frac{d_x}{2} \\ \frac{d_y}{2} \end{pmatrix}$ belongs to the image of the last diaphragm.

The equation (7) defines the apparent image of the last diaphragm on the mirror image. It is the equation of an ellipse, with the same axis as the curvature tensor $\vec{C}_\Sigma$ and with its semi-major (minor) axis respectively proportional to R$_1$ and R$_2$.

$$R_1^{mirror} = \frac{R_1\Phi}{2H} \tag{8}$$

and

$$R_2^{mirror} = \frac{R_2\Phi}{2H} \tag{9}$$

In the case of a point charge, it has been demonstrated by another way by Vallayer & al that the radius R of the equipotential corresponding to the electron trajectory that impacts the edge of the output diaphragm is linked to the geometric features of the experiment by the relation [5]

$$R = \frac{d_{mirror}}{\Phi}H \tag{10}$$

In that case, the mirror image is a circle and not an ellipse. The equation (10) is completely coherent with the equations (7), (8) and (9) for a point charge with R$_1$=R$_2$=R.

As a partial conclusion, when considering an equipotential of any shape, the last diaphragm mirror image obtained by the reflection of a lecture electron beam is an ellipse whose features are directly linked to the curvature tensor of the equipotential (equation 7). This curvature tensor depends on the potential lecture, ie on the kinetic energy of the electron. But this link has to be determined for each trapped charge distribution and for isotropic or anisotropic material.

*2.4 Mirror characteristic in the axisymetric case*
For any axisymetric charge distribution, the potential field near the impact point M$_0$ can be written (figure. 2b):

$$V(z,x) = V_0(z) + \frac{1}{2}x^2V_2(z) + \frac{1}{4!}x^4V_4(z) + ... \tag{11}$$

The potential has to follow the Laplace equation $\Delta V = 0$ that leads to

$$V_2 = -\frac{\partial^2 V_0}{\partial z^2} \text{ and } V_4 = -\frac{\partial^4 V_0}{\partial z^4} \qquad (12)$$

Moreover, locally i.e. near the point $z = z_0$ on the axis, one can write

$$\left(\frac{\partial V_0}{\partial z}\right)_0 (z-z_0) + \frac{1}{2}x^2 V_2(z_0) = 0 \qquad (13)$$

This corresponds to the local equation of a circle

$$x^2 + (R_{axisym} - (z-z_0))^2 = R^2_{axisym} \qquad (14)$$

with radius

$$R_{axisym} = \frac{2\left(\frac{\partial V_0}{\partial z}\right)}{\left(\frac{\partial^2 V_0}{\partial z^2}\right)} \qquad (15)$$

So, in the case of an axisymetric distribution of trapped charges, the curvature tensor radii of the equipotentials are equal to $R_{axisym}$. By replacing this value in the general equation (7), one obtains the corresponding mirror image radius in the axisymetric case:

$$R_{mirror-axisym} = \frac{\phi}{H} \frac{\left(\frac{\partial V}{\partial z}\right)}{\left(\frac{\partial^2 V}{\partial z^2}\right)} \qquad (16)$$

This expression obtained from the general equation (7), in the case of axisymetric charge distribution, is the same as the one obtained by other authors [9].

### 3. Anisotropic distribution of charges in a linear homogeneous isotropic dielectric (LHI where $\varepsilon_x = \varepsilon_y = \varepsilon_z = \varepsilon_r$)

Usually, trapped charge distributions are considered as isotropic and the classical distribution shape used in the models is the sphere. However it is interesting to determine the potential and the corresponding mirror formula in the case of anisotropic distribution. In this case, in the vacuum or in a LHI, the homeoidal distribution generalizes the spherical distribution to the anisotropic case. The homeoidal distribution corresponds to a uniform distribution of charge between two infinitely neighbouring homothetic ellipsoids. It has been demonstrated for example by Durand, that this homeoidal distribution is the solution in the case of a conductive ellipsoid [10]. The equipotentials $V(\xi)$ are then confocal ellipsoids which can be defined by the following equations:

$$\frac{x^2}{a^2+\xi} + \frac{y^2}{b^2+\xi} + \frac{z^2}{c^2+\xi} = 1 \qquad (17)$$

and the associated matrices M are :

$$M = \begin{pmatrix} \frac{1}{a^2+\xi} & 0 & 0 \\ 0 & \frac{1}{b^2+\xi} & 0 \\ 0 & 0 & \frac{1}{c^2+\xi} \end{pmatrix} \qquad (18)$$

The potential V(ξ) created by the homeoidal distribution can be found in Durand's book and is given by

$$V(\xi) = \frac{Q}{4\pi\varepsilon_0(\varepsilon_r+1)} \int_\xi^\infty \frac{d\lambda}{R_\lambda} \quad \text{with} \quad R_\lambda = \sqrt{(\lambda+a^2)(\lambda+b^2)(\lambda+c^2)} \tag{19}$$

Thanks to the equation (19), it will be possible to determine the relation between the features of the ellipsoid (a, b, c) versus the potential V(ξ) and so versus the kinetic energy of the incident electrons. Moreover as demonstrated in Appendix A equation (A.6), at a point $M_0$ on the z axis, the curvature tensor is directly linked to the matrix M:

$$C_\Sigma = \frac{M_\Sigma}{\|M.\vec{x}_0\|} = \sqrt{c^2+\xi} \begin{pmatrix} \frac{1}{a^2+\xi} & 0 \\ 0 & \frac{1}{b^2+\xi} \end{pmatrix} \tag{20}$$

By combining the equations (7), (19) and (20) the mirror image equation can be established for different limit cases.

*3.1. Trapped charge distributed on a segment [-a, a] oriented along the x axis.*
One considers a segment [-a, a] of trapped charges (figure 3(a)) oriented along the x axis in the (Oxy) plane of electron injection.

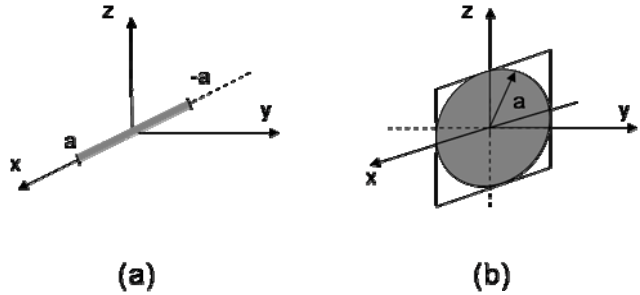

(a) (b)

**Figure 3.** Two limit cases of the homeoidal charge distribution (a) uniformly charge segment [-a, a] oriented along the x axis (b) homeoidal charge distributed on a disk of radius a in the Oxz plane. Electron injection direction is along the z-axis.

The matrix $M_{segment}$ of the equipotential V(ξ) corresponding to this homeoidal distribution on a segment is the limit case of the homeoidal distribution with b and c tending towards zero:

$$M_{segment} = \begin{pmatrix} \frac{1}{a^2+\xi} & 0 & 0 \\ 0 & \frac{1}{\xi} & 0 \\ 0 & 0 & \frac{1}{\xi} \end{pmatrix} \tag{21}$$

Note that, since $\|M.\vec{x}_0\| = 1/\sqrt{\xi}$ for $M_0(0,0,z)$ belonging to Σ (so that $z = \sqrt{\xi}$), the corresponding curvature tensor is then

$$C^{segment}_\Sigma = \begin{pmatrix} \frac{1}{a^2+\xi}\sqrt{\xi} & 0 \\ 0 & \frac{1}{\sqrt{\xi}} \end{pmatrix} \tag{22}$$

or, more precisely:

$$\frac{1}{R_x} = \frac{1}{(a^2+\xi)}\sqrt{\xi} \tag{23}$$

$$\frac{1}{R_y} = \frac{1}{\sqrt{\xi}} \tag{24}$$

From equation (19), the corresponding potential is then given by:

$$V(\xi) = \frac{Q}{4\pi\varepsilon_0(\varepsilon_r+1)} \int_\xi^\infty \frac{d\lambda}{\lambda\sqrt{(\lambda+a^2)}} \tag{25}$$

Assume we have a change of variable $\lambda = a^2 sh^2(y)$, one obtains [11]:

$$V(\xi) = \frac{Q}{2\pi\varepsilon_0(\varepsilon_r+1)}\frac{1}{a}\int_\eta^\infty \frac{dy}{sh(y)} = \frac{Q}{2\pi\varepsilon_0(\varepsilon_r+1)}\frac{1}{a}\operatorname{arg\,coth}(ch\eta) \tag{26}$$

with $\xi = a^2 sh^2(\eta)$. From equation (26), one deduces $ch(\eta) = \coth(\frac{2\pi\varepsilon_0(\varepsilon_r+1)V(\xi)a}{Q})$.

Finally, by replacing $\xi$ in equations (23), (24) by its value, and by taking into account the equations (8) and (9), the image of the last output diaphragm is given by:

$$\frac{1}{R_x^{mirror}} = \frac{2H}{a\Phi}\frac{sh(\frac{2\pi\varepsilon_0(\varepsilon_r+1)2V_la}{Q})}{ch^2(\frac{2\pi\varepsilon_0(\varepsilon_r+1)2V_la}{Q})} \tag{27}$$

$$\frac{1}{R_y^{mirror}} = \frac{2H}{a\Phi}sh(\frac{2\pi\varepsilon_0(\varepsilon_r+1)2V_la}{Q}) \tag{28}$$

In the case of a charge Q homogeneously distributed on a linear segment [-a, a] along the Ox axis, the mirror image is elliptic, with $R_x^{mirror}$ and $R_y^{mirror}$ the main-major (respectively minor) axis defined just above.

Note also that in the case of a segment perpendicular to the electron injection plane (along Oz), the mirror image is circular, and $1/R_x^{mirror} = 1/R_y^{mirror} = \frac{H}{a\Phi}sh(\frac{2\pi\varepsilon_0(\varepsilon_r+1)2V_la}{Q})$. We find for this isotropic case the same results as Ghorbel et al [9].

*3.2. Trapped charge distributed on a disk of radius a included in the plane Oxz.*

Another interesting limit case of the anisotropic homeoidal distribution of trapped charges is a disk of radius a (figure 3(b)). Note that for such homeoidal charge on a disk, the charge density is proportional to $(a^2-r^2)^{-1/2}$ at the distance r of the disk center [10]. This disk is for example included in the Oxz plane, parallel to the electron injection direction and perpendicular to the y axis. The matrix M of the equipotential $V(\xi)$ corresponding to this homeoidal distribution on a disk is the limit case of the homeoidal distribution with a=c and b tending towards zero:

$$M_{disk} = \begin{pmatrix} \frac{1}{a^2+\xi} & 0 & 0 \\ 0 & \frac{1}{\xi} & 0 \\ 0 & 0 & \frac{1}{a^2+\xi} \end{pmatrix} \tag{29}$$

and, since $\|M.\vec{x}_0\| = 1/\sqrt{a^2+\xi}$ for $M_0(0,0,z)$ belonging to $\Sigma$ (so that $z = \sqrt{a^2+\xi}$), the corresponding curvature tensor is

$$C^{disk}_\Sigma = \sqrt{a^2+\xi} \begin{pmatrix} \frac{1}{a^2+\xi} & 0 \\ 0 & \frac{1}{\xi} \end{pmatrix} \qquad (30)$$

From equation (19), for the disk, the corresponding potential is:

$$V(\xi) = \frac{Q}{4\pi\varepsilon_0(\varepsilon_r+1)} \int_\xi^\infty \frac{d\lambda}{(\lambda+a^2)\sqrt{\lambda}} \qquad (31)$$

Assume we have the same change of variable $\lambda = a^2 sh^2(y)$, one obtains (Gradshteyn, 1965 #1055):

$$V(\xi) = \frac{Q}{2\pi\varepsilon_0(\varepsilon_r+1)} \frac{1}{a} \int_\eta^\infty \frac{dy}{ch(y)} = \frac{Q}{2\pi\varepsilon_0(\varepsilon_r+1)} \frac{1}{a} \operatorname{arc cot g}(sh\eta) \qquad (32)$$

with $\xi = a^2 sh^2(\eta)$. From equation (32), one deduces $sh(\eta) = \cot g(\frac{2\pi\varepsilon_0(\varepsilon_r+1)V(\xi)a}{Q})$.

Then, by using equations (7), (8) and (9) as previously, one obtains the curvature radii $R_x$ and $R_y$ of the equipotential and the ellipse main-axis $R_x^{mirror}$ and $R_y^{mirror}$ of the last output diaphragm mirror image.

$$\frac{1}{R_x^{mirror}} = \frac{2H}{a\Phi} \frac{\sin(\frac{2\pi\varepsilon_0(\varepsilon_r+1)2V_l a}{Q})}{\cos^2(\frac{2\pi\varepsilon_0(\varepsilon_r+1)2V_l a}{Q})} \qquad (33)$$

$$\frac{1}{R_y^{mirror}} = \frac{1}{a} \sin(\frac{2\pi\varepsilon_0(\varepsilon_r+1)2V_l a}{Q}) \qquad (34)$$

In the case of a homeoidal charge Q distributed on a disk perpendicular to the plane of the electron injection, the mirror image is elliptic, with $R_x^{mirror}$ and $R_y^{mirror}$ the main-major (respectively minor) axis defined just above.

## 4. Screening of a charge lying at the interface between an anisotropic orthotropic dielectric and vacuum

In the previous paragraphs, the case of LHI dielectrics has been considered and the characteristics of the equipotentials induced by the presence of an anisotropic charge distribution in such materials have been demonstrated. Furthermore, the features of the mirror image obtained after an electron injection of such anisotropic distribution in an isotropic material have been also evaluated. This paragraph is devoted to the case of linear homogeneous orthotropic anisotropic dielectrics (LHOA). Indeed these materials are numerous such as titanium dioxide $TiO_2$, quartz, etc, but also such as new synthetic nanomaterials, whose dielectric properties can also be anisotropic such as copolymers, filled materials...
In all the following, the considered dielectrics are orthotropic materials with possibly different dielectric properties in all three different orthogonal directions (x,y,z) and with an interface compatible with the plane of electron injection (defined as z = 0). That means that the interface between the two dielectrics corresponds to the plane of electron injection and that one of the principal axes of the permittivity tensor is orthogonal to the injection plane and the two others are in that plane.
Prior to solve the problem of a charge lying at the interface between the vacuum and a LHOA dielectric, the transformation of various solutions of the electrostatic equations by different simple geometrical transformations will be first considered.

*4.1. Modification of the electrostatic solutions after an affine transformation*

Consider a linear transformation A that transforms the initial coordinates system (ICS) (x,y,z) in the transformed coordinates system (TCS), (X,Y,Z) = A (x,y,z) where A is a 3x3 matrix. One notes the gradient (nabla) operator in the TCS system: $\widetilde{\vec{\nabla}} = \frac{\partial}{\partial \vec{X}} = A^{-1}.\frac{\partial}{\partial_x} = A^{-1}.\vec{\nabla}$.

Let φ(x,y,z) = Cte be the equipotential surfaces network in the ICS where the dielectric permittivity is [ε], and that corresponds to a charge density ρ(x,y,z) related to φ by the Maxwell-Poisson law. Two cases can be considered concerning the transformation of the charge density and the permittivity. Indeed the Maxwell-Poisson law in the TCS can be written $\widetilde{\rho}(X,Y,Z) = -\widetilde{\vec{\nabla}}.([\widetilde{\varepsilon}].\widetilde{\vec{\nabla}}\widetilde{\varphi}(X,Y,Z))$, so that the network of equipotential φ(x,y,z) = Cte can be transformed in the corresponding network of equipotential $\widetilde{\varphi}(X,Y,Z)$ = φ(x,y,z) for a new charge distribution else (a) by transforming the dielectric constant and keeping the density of charges constant, or (b) by changing the density of charges and keeping the permittivity constant.

In the first case (a), one considers the affinity $A = \begin{pmatrix} a_1 & 0 & 0 \\ 0 & a_2 & 0 \\ 0 & 0 & a_3 \end{pmatrix}$, the permittivity of the new material becomes $[\widetilde{\varepsilon}] = {}^t A.[\varepsilon].A$, the density of charges is unchanged $\widetilde{\rho}(X) = \rho(x)$ but the charges in a transformed volume are modified because of the non-conservative volume $dQ = |\det A| dq$.

In the second case (b), one considers an homothetic transformation $H = \begin{pmatrix} k & 0 & 0 \\ 0 & k & 0 \\ 0 & 0 & k \end{pmatrix}$ with a dilatation factor k, in this transformation the permittivity is kept constant $[\widetilde{\varepsilon}] = [\varepsilon]$, but the volume charge is changed according to the Maxwell-Poisson law $\widetilde{\rho}(X) = \frac{1}{k^2} \rho(x)$ and $dQ = k dq$.

*4.2. Example of application to the case of a point charge in an isotropic dielectric*

This method of linear transformations can be applied to find the classical formula of electrostatic image charge in the case of a point charge Q leading at the interface between two isotropic materials with respective permittivity $\varepsilon_1$ and $\varepsilon_2$. This simple case is useful to understand how this method works, and is summarized in the figure 4. We start from a configuration which potential solution is known: a point charge q in an LHI (permittivity $\varepsilon_1$). To solve the singularity, the point charge is assimilated to a spherically spread charge q, composed of a charge q/2 in the upper half-space and also of a q/2 charge in the lower half-space. This configuration is locally equivalent to the point charge distribution and fulfils the potential continuity at the interface. The external potential is then given by:

$$V(r) = \frac{q}{4\pi\varepsilon_0\varepsilon_1 r} \quad (35)$$

In the first step, an affine transformation corresponding to the previous case (a), which matrix A is $A = \begin{pmatrix} k & 0 & 0 \\ 0 & k & 0 \\ 0 & 0 & k \end{pmatrix}$ with $k = \sqrt{\frac{\varepsilon_2}{\varepsilon_1}}$ is applied to the lower half-space.

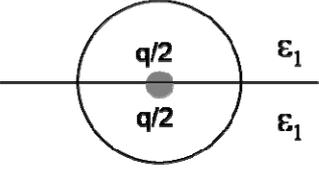

**Figure 4**. Successive linear transformations used to find the potential value induced by a charge located at the interface between two isotropic dielectrics, starting from the known problem of a charge embedded in an isotropic insulator.

After this first transformation, the ICS (x,y,z) is transformed in the TCS(1) $(X_1,Y_1,Z_1)$ and a new equipotential network is obtained $\widetilde{\varphi}(X_1,Y_1,Z_1) = \varphi(x,y,z)$ with the right permittivities $\varepsilon_1$ and $\varepsilon_2$ in the two half-spaces but the condition of the continuity of the potential is not fulfilled at the interface. So, in a second step, a homothety H is applied with the dilatation factor 1/k. After this second transformation, the wanted problem of a point charge $Q = \frac{q}{2} + \frac{Q''}{2} = \frac{q}{2}(1+k^2)$ at the interface between two different LHI with respective permittivities $\varepsilon_1$ and $\varepsilon_2$ is obtained and the corresponding potential is given by

$$V(r) = \frac{q}{4\pi\varepsilon_0\varepsilon_1 r} = \frac{2Q}{4\pi\varepsilon_0\varepsilon_1(1+k^2)r} = \frac{1}{4\pi\varepsilon_0 r}\frac{2Q}{(\varepsilon_1+\varepsilon_2)} \qquad (36)$$

With this simple method of applying successive linear transformations compatible with the Maxwell-Poisson law, the classical potential of a charge lying at the interface of two different LHI dielectrics is found again [10], [12].

We will now use this method in the case of a point charge lying at the interface between the vacuum and an orthotropic anisotropic dielectrics (LHOA, where $\varepsilon_x \neq \varepsilon_y \neq \varepsilon_z$). This case has been soon solved by Mele by another way in the case of a linear homogeneous isotropic-transverse dielectric (LHIT, where $\varepsilon_x = \varepsilon_y \neq \varepsilon_z$) but not in the more general case of the LHOA [12]. By transforming the reference solution of the previously solved problem (of a charge lying at the interface of two different LHI), one can generate solutions corresponding to other couple of dielectrics thanks to successive linear transformations. However one condition is needed: the restriction of the different linear transformations $A_i$ to the interface plane $\Sigma$, ($A_{i\Sigma}$) must differed by simple homotheties, so that charges and equipotentials can be changed to recover the continuity of the potential at the interface $\Sigma$.

*4.3. Determination of the equivalent charge and potential for a charge at a vacuum/dielectric interface in the case of an orthotropic anisotropic dielectric*

*4.3.1. Reference electrostatic problem.* Let first consider the case of a point charge q at the interface between two LHOA dielectrics with permittivities $\varepsilon_{1ref}$ and $\varepsilon_{2ref}$ defined as follow:

$$\varepsilon_{1ref} = \begin{pmatrix} \alpha & 0 & 0 \\ 0 & 1/\alpha & 0 \\ 0 & 0 & \beta \end{pmatrix} \text{ and } \varepsilon_{2ref} = \begin{pmatrix} 1/\alpha & 0 & 0 \\ 0 & \alpha & 0 \\ 0 & 0 & \beta \end{pmatrix} \text{ with } \beta = \frac{1}{2}\left(\alpha + \frac{1}{\alpha}\right)$$

The equipotentials for such a problem can be approximated in a first time by spheres (that imposed the value of $\beta$ versus $\alpha$), and then the potential can be written:

$$V(r) = \frac{Cq}{4\pi\varepsilon_0 r} \tag{37}$$

where $C=1/\beta$ is a constant, that is determined by solving the Maxwell-Poisson equation and by fulfilling the potential continuity equation. This evaluation and the choice of $\beta$ are detailed in Appendix B. One obtains from equation (B.4):

$$V(r) = \frac{q}{4\pi\varepsilon_0 r}\frac{1}{\beta} \tag{38}$$

*4.3.2. General case of anisotropic (LHOA) dielectrics*

To obtain the approximate expression of the field created by a point charge located at the interface between the vacuum and a dielectric which permittivity is anisotropic, the following method is used and summing up in figure 5.

(i) We start from a reference problem that has been previously solved (§ 3.3.1) of a point charge lying at the interface between two LHOA dielectrics with permittivities $\varepsilon_{1ref}$ and $\varepsilon_{2ref}$, as defined previously. The approximate solution is supposed to be given by an expression with spherical symmetry (cf. Appendix B).

(ii) We apply the right affinities $A_1$ and $A_2$ at constant $\rho$ (as defined in case (a) of § 4-1) to each half-space so that the permittivities take the expected values respectively $\begin{pmatrix} 1 & 0 & 0 \\ 0 & 1 & 0 \\ 0 & 0 & 1 \end{pmatrix}$ for the vacuum in the upper half-space, and $\begin{pmatrix} \varepsilon_x & 0 & 0 \\ 0 & \varepsilon_y & 0 \\ 0 & 0 & \varepsilon_z \end{pmatrix}$ for the LHOA dielectric in the lower half-space.

One obtains $\varepsilon'_1 = {}^tA_1 \cdot \varepsilon_1 \cdot A_1$, $a_1 = \sqrt{\varepsilon_x \alpha}$, $b_1 = \sqrt{\dfrac{\varepsilon_y}{\alpha}}$ and $c_1 = \sqrt{\dfrac{\varepsilon_z}{\beta}}$ for the lower half space and for the upper half-space $\varepsilon'_2 = {}^tA_2 \cdot \varepsilon_2 \cdot A_2$, i.e. $a_2 = \dfrac{1}{\sqrt{\alpha}}; b_2 = \sqrt{\alpha}; c_2 = \dfrac{1}{\sqrt{\beta}}$.

| Permittivity | Equipotentials | Charge | Charge | Equipotentials | Permittivity |
|---|---|---|---|---|---|
| Lower half space | | | Upper half space | | |
| $\varepsilon_1 = \begin{pmatrix} 1/\alpha & 0 & 0 \\ 0 & \alpha & 0 \\ 0 & 0 & \beta \end{pmatrix}$ | $x^2 + y^2 + z^2 = 1$ | $\dfrac{q}{2}$ | $\dfrac{q}{2}$ | $x^2 + y^2 + z^2 = 1$ | $\varepsilon_2 = \begin{pmatrix} \alpha & 0 & 0 \\ 0 & 1/\alpha & 0 \\ 0 & 0 & \beta \end{pmatrix}$ |
| $A_1 = \begin{pmatrix} a_1 & 0 & 0 \\ 0 & b_1 & 0 \\ 0 & 0 & c_1 \end{pmatrix}$ | Affine transformations $A_1$ and $A_2$ allowing to obtain the expected permittivities $\varepsilon'_1$ and $\varepsilon'_2$, at constant $\rho$ | | | | $A_2 = \begin{pmatrix} a_2 & 0 & 0 \\ 0 & b_2 & 0 \\ 0 & 0 & c_2 \end{pmatrix}$ |
| $\varepsilon'_1 = \begin{pmatrix} \varepsilon_x & 0 & 0 \\ 0 & \varepsilon_y & 0 \\ 0 & 0 & \varepsilon_z \end{pmatrix}$ $\varepsilon'_1 = {}^tA_1 \cdot \varepsilon_1 \cdot A_1$ $\varepsilon_x = \dfrac{a_1^2}{\alpha}; \varepsilon_y = b_1^2 \alpha;$ $\varepsilon_z = c_1^2 \beta$ | ${}^tX \cdot A_1^2 \cdot X = 1$ | $\det A_1 \dfrac{q}{2}$ | $\det A_2 \dfrac{q}{2}$ | ${}^tX \cdot A_2^2 \cdot X = 1$ | $\varepsilon'_2 = \begin{pmatrix} 1 & 0 & 0 \\ 0 & 1 & 0 \\ 0 & 0 & 1 \end{pmatrix}$ $\varepsilon'_2 = {}^tA_2 \cdot \varepsilon_2 \cdot A_2$ $a_2 = \dfrac{1}{\sqrt{\alpha}}, b_2 = \sqrt{\alpha};$ $c_2 = \dfrac{1}{\sqrt{\beta}}$ |
| $H_1 = \begin{pmatrix} k & 0 & 0 \\ 0 & k & 0 \\ 0 & 0 & k \end{pmatrix}$ | Homothety $H_1$ at constant $\varepsilon$ with a dilatation factor $k$ applied on the lower half space to fulfil the continuity condition of the potential at the interface | | | | |
| $\varepsilon'_1 = \begin{pmatrix} \varepsilon_x & 0 & 0 \\ 0 & \varepsilon_y & 0 \\ 0 & 0 & \varepsilon_z \end{pmatrix}$ | ${}^tX \cdot k^2 A_1^2 \cdot X = 1$ | $k \det A_1 \dfrac{q}{2}$ | $\det A_2 \dfrac{q}{2}$ | ${}^tX \cdot A_2^2 \cdot X = 1$ | $\varepsilon'_2 = \begin{pmatrix} 1 & 0 & 0 \\ 0 & 1 & 0 \\ 0 & 0 & 1 \end{pmatrix}$ |
| Continuity condition at the interface: ${}^tXk^2A_1^2 \cdot X = {}^tX \cdot A_2^2 \cdot X \quad \forall X = (X, Y, 0)$ $\begin{cases} k^2 a_1^2 = a_2^2 \\ k^2 b_1^2 = b_2^2 \end{cases}$ $\begin{cases} k^4 = \dfrac{1}{\varepsilon_x \varepsilon_y} \\ \alpha^4 = \dfrac{\varepsilon_y}{\varepsilon_x} \end{cases}$ | | | | | |

**Figure 5.** Successive linear transformations used to find the potential value and the equivalent charge induced by a charge located at the interface between vacuum and an anisotropic dielectric (LHOA), starting from the known problem of a charge at the interface between two related dielectrics.

(iii) A homothety $H_1$ at constant permittivity (as defined in case (b) of § 4.1) with a dilatation factor k is applied to the lower half-space so that the equipotentials of the two media correspond. Note that these equipotentials are ellipsoids different in each half-space, and the homothety is used to recover the potential continuity. This imposes that α and k can not be chosen but that their values are fixed and one obtains $\alpha = \sqrt[4]{\frac{\varepsilon_y}{\varepsilon_x}}$, $\beta = \frac{1}{2}\left(\sqrt[4]{\frac{\varepsilon_y}{\varepsilon_x}} + \sqrt[4]{\frac{\varepsilon_x}{\varepsilon_y}}\right)$ and $k = \sqrt[4]{\frac{1}{\varepsilon_x \varepsilon_y}}$.

Finally, the approximate solution of the potential created by a point charge located at the interface between the vacuum and a LHOA dielectric is obtained and the charge Q that creates the field can also be deduced from the charge q of the reference problem:

$$Q = \frac{q}{2}\det A_2 + \frac{k\,q}{2}\det A_1 \qquad (39)$$

and by replacing the different terms by their values in function of $\varepsilon_x$, $\varepsilon_y$, $\varepsilon_z$, one obtains the equivalent charge seen from the vacuum:

$$Q = \frac{q}{2}\frac{1}{\sqrt{\beta}}\left(1 + \sqrt{\varepsilon_z \sqrt{\varepsilon_x \varepsilon_y}}\right) \quad \text{with} \quad \beta = \frac{1}{2}\left(\sqrt[4]{\frac{\varepsilon_y}{\varepsilon_x}} + \sqrt[4]{\frac{\varepsilon_x}{\varepsilon_y}}\right) \qquad (40)$$

Also, seen from the vacuum and taking into account the change of the coordinate system $(X,Y,Z) = A_2 (x,y,z)$ with $r = \sqrt{\left(x^2 + y^2 + z^2\right)} = \sqrt{\left(\alpha X^2 + \frac{Y^2}{\alpha} + \beta Z^2\right)}$ and $\alpha = \sqrt[4]{\frac{\varepsilon_y}{\varepsilon_x}}$, the potential created by the charge Q, which follows the equation (39), is given by $\widetilde{\varphi}(X,Y,Z) = \varphi(x,y,z) = \frac{q}{4\pi\varepsilon_0 r}\frac{1}{\beta}$. By replacing q and β by their respective values, one obtains:

$$\widetilde{\varphi}(X,Y,Z) = \frac{1}{4\pi\varepsilon_0}\frac{2Q}{\left(1 + \sqrt{\varepsilon_z \sqrt{\varepsilon_x \varepsilon_y}}\right)}\sqrt{\frac{2}{\alpha + \frac{1}{\alpha}}}\frac{1}{\sqrt{\left(\alpha X^2 + \frac{Y^2}{\alpha} + \beta Z^2\right)}} \qquad (41)$$

with $\alpha = \sqrt[4]{\frac{\varepsilon_y}{\varepsilon_x}}$ and $\beta = \frac{1}{2}\left(\sqrt[4]{\frac{\varepsilon_y}{\varepsilon_x}} + \sqrt[4]{\frac{\varepsilon_x}{\varepsilon_y}}\right)$.

The equation (41) gives the potential in the vacuum, created by a charge Q located at the interface between the vacuum and a linear homogeneous orthotropic anisotropic material.

If this material is a isotropic-transverse dielectric (LHIT) i.e. such that $\varepsilon_x = \varepsilon_y = \varepsilon_{IT}$ then α=β=1 and equation (41) becomes

$$\varphi_{IT}(X,Y,Z) = \frac{1}{4\pi\varepsilon_0}\frac{2Q}{\left(1 + \sqrt{\varepsilon_z \varepsilon_{IT}}\right)}\frac{1}{\sqrt{\left(X^2 + Y^2 + Z^2\right)}} \qquad (42)$$

The same result as the one published by Mele are obtained for this particular case of transversal isotropic dielectrics (eq (32) in [12] with s=0).

Note that for a transversal isotropic material, when the charge is lying at the interface or inside the material, the equipotentials in the upper half-space are circular. The same method of linear transformation can also be applied for charges inside the dielectric but the corresponding calculations will not be presented here (to not increase too much the length of this article). The obtained result is that, for a charge located inside the dielectric at a distance s, the anisotropy will induce an apparent displacement of the charge at the position s' for an observer situated in the vacuum, and the charge has to be renormalized.

$$s' = \sqrt{\frac{\varepsilon_{IT}}{\varepsilon_z}} d \tag{43}$$

*4.4. Comparison with the equipotentials obtained by finite element simulation*

The previous analytical calculations (equation (41)) of the equipotential network created by a point charge lying at the interface between the vacuum and a LHOA dielectric give the following equation of the equipotentials:

$$\left( \alpha X^2 + \frac{Y^2}{\alpha} + \beta Z^2 \right) = C \tag{44}$$

The obtained equipotential are then ellipsoids with the three respective values for the main-axes:

$a \propto \frac{1}{\sqrt{\alpha}}; b \propto \sqrt{\alpha}; c \propto \frac{1}{\sqrt{\beta}}$ with $\alpha = \sqrt[4]{\frac{\varepsilon_y}{\varepsilon_x}}$ and $\beta = \frac{1}{2}\left( \sqrt[4]{\frac{\varepsilon_y}{\varepsilon_x}} + \sqrt[4]{\frac{\varepsilon_x}{\varepsilon_y}} \right)$. The intersections of these ellipsoids with the Z=0 plane, are ellipses whose ratios of the corresponding main-axis values a/b is equal to $1/\alpha$.

By using the finite element simulation program COMSOL Multiphysics™, we can simulate the presence of a quasi-point charge (100 pC located in a hemi-sphere of radius 5µm) at the interface between the vacuum and a LHOA dielectric. Figure 6 gives the obtained equipotentials in the case of a LHOA material with $1/\alpha \approx 1.19$ i.e. with the same dielectric properties as TiO2 rutile. The direction of electron injection and the values of the permittivities are given on the figure 6. The calculated equipotentials far away from the charge (to keep the approximation of a point charge) are effectively ellipsoids. The ratios of their respective main axes value in the potential range $300V < V_\xi = 2V_1 < 600V$, that corresponds approximately to the experimental conditions where the point charge approximation is valid, vary in the range [1.1, 1.3] and are approximately equal to the expected value $1/\alpha \approx 1.2$ for a low lecture potential but this ratio increases for higher potential (near the stored charge).

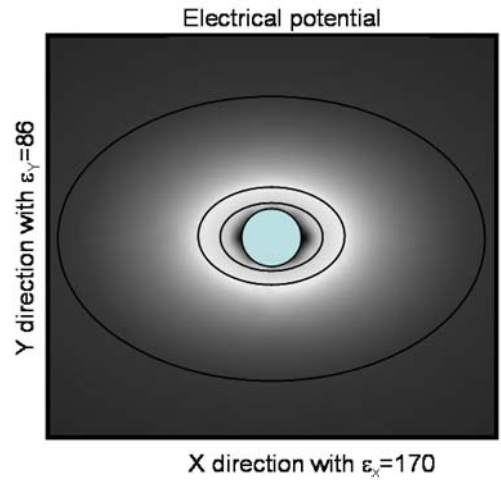

**Figure 6.** Projection of the equipotential ellipsoids onto the $OX_2Y_2$ plane obtained by COMSOL™ finite element simulation in the case of a hemispherical charge located at the interface between the vacuum and TiO$_2$, for low potential values. The interface plane is the (110) plane, the permittivities in the OXY plane are respectively $\varepsilon_X$=86 $\varepsilon_Y$=170.

*4.5. Consequence of the elliptic equipotentials on the mirror image*

Returning to the general case of LHOA dielectrics, one can determine from equation (41), the equipotentials created by a point charge located at the interface between the vacuum and a LHOA dielectric. They are given by $\left(\alpha X^2 + \dfrac{Y^2}{\alpha} + \beta Z^2\right) = C$, with C a constant. Their associated matrices are then $M = \begin{pmatrix} \dfrac{\alpha}{C} & 0 & 0 \\ 0 & \dfrac{1}{\alpha C} & 0 \\ 0 & 0 & \dfrac{\beta}{C} \end{pmatrix}$, by applying the results obtained in Appendix A, the corresponding curvature tensor is given by $C_\Sigma = \dfrac{M_\Sigma}{\|M.\vec{x}_0\|}$, in $M_0(0,0,Z)$, belonging to the equipotential $\beta Z^2 = C$ and $\|M.\vec{x}_0\| = \dfrac{\beta}{C} Z = \sqrt{\dfrac{\beta}{C}}$, so the curvature radii of the equipotential on the e-beam axis are equal to:

$$\frac{1}{R_x} = \frac{\alpha}{\sqrt{C\beta}} \quad \text{and} \quad \frac{1}{R_y} = \frac{C}{\alpha\sqrt{\beta}} \tag{42}$$

From equations (8), (9) and (42), one can deduce (i) that if $\varepsilon_x \neq \varepsilon_y$, then the mirror image created by a point charge stored at the interface between the vacuum and an anisotropic material is elliptic and not circular like in the case of a point charge in a isotropic dielectric, (ii) that the ratio between the two main axes values of the elliptic mirror image is given by

$$\frac{R_x}{R_y} = \frac{1}{\alpha^2} = \sqrt{\frac{\varepsilon_x}{\varepsilon_y}} \tag{43}$$

This equation demonstrates that for a point charge at the interface between vacuum and an anisotropic material (not LHIT), the major axis of the induced elliptic mirror image is along the direction of higher permittivity.
Note also that the variation of the two main axes values $R_x$ and $R_y$ of the elliptic mirror versus the lecture potential $V_l$ is such that their ratio is kept constant as far as the point charge approximation is valid, and equal to the value given in equation (43).

## 5. Example of experimental mirror curves on TiO2

*5.1. Material and associated coordinate systems*
Titanium dioxide presents three principal crystallographic structures: rutile, anatase and brookite. Rutile is the common and most stable form. In this study, oriented (110) and (001) samples of pure single-crystal rutile were used. Rutile crystallizes in a tetragonal system whose lattice parameters are a = 0.495 nm and c = 0.259 nm. As shown in our previous publication, rutile traps electric charges only at low temperature, due to strong leakage current and too weak traps [7].
If one takes a first coordinate system $(X_1, Y_1, Z_1) = $ (CS1) wherein axes are respectively along the [110], [1$\bar{1}$0], and [001] directions of the crystallographic coordinate system, the permittivity tensor is diagonalized. This CS1 is used for the electron injection performed in the [001] direction, for which the interface is the plane (001), so that the rutile can be considered as a linear homogeneous transverse-isotropic material in this configuration. Indeed, in this coordinate system (CS1), the anisotropy of the relative dielectric constant of rutile is characterized at 300K by the following dielectric tensor [13] and $\varepsilon_X = \varepsilon_Y$:

$$\varepsilon_{\text{rutile}/(\text{CS1})} = \begin{pmatrix} 86 & 0 & 0 \\ 0 & 86 & 0 \\ 0 & 0 & 170 \end{pmatrix} \tag{45}$$

For the electron injection in the direction [110], a second coordinate system (CS2) has been chosen, wherein axes are respectively along the [001], [1$\bar{1}$0], and [110] directions. In that particular orientation of the crystal compared to the electron injection, the dielectric permittivity is not axisymetric versus the electron injection direction, indeed $\varepsilon_X$=170 and $\varepsilon_Y$=86. In the CS2, the dielectric tensor is written:

$$\varepsilon_{\text{rutile}/(CS2)} = \begin{pmatrix} 170 & 0 & 0 \\ 0 & 86 & 0 \\ 0 & 0 & 86 \end{pmatrix} \quad (46)$$

*5.2. Comparison between experiment and calculation.*
Figure 7 shows the mirror images obtained for the two directions of injection, at relatively high lecture potential. The mirror shape is circular in the case of (001) plane (figure 7(a)) and elliptic in the case of (110) plane (figure 7(b)). This anisotropy of the mirror image can be due to two simultaneous effects: (i) to the anisotropy of the trapped charge distribution itself, possibly due to the anisotropy of the material (ii) to the anisotropy of the dielectric permittivity even if the stored charge distribution is isotropic.

*5.2.1. Electron injection in the [001] direction.* The fact that the mirror is circular, in the case of [001] injection direction, is consistent with the above analytical calculations (eq.(43), (44)). Indeed, we showed that for this direction of injection, rutile can be considered as tranverse-isotropic, and $\alpha$ is equal to 1. Thus, the radii $R_x$ and $R_y$ of the curvature tensor, and thus of the mirror image, are equals when the point charge approximation is verified. At higher lecture potential, the mirror image remains circular. From our calculations in part III, we showed that if the charge distribution is not axisymetric compared to the injection direction the image mirror becomes elliptic. We can then conclude, from the fact that the mirror image stays circular, that the trapped charge distribution is axisymetric.

*5.2.2. Electron injection in the [110] direction.* Experimentally an elliptic mirror image is obtained in this direction of injection as seen in figure 7 (b). When the sample is rotated in the (110) plane, the elliptic mirror image is also rotated of the same angle, confirming the crystallographic nature of the observed anisotropy. As predicted by our previous calculations the major axis of the mirror is along the [001]=$X_2$ direction that is the direction of highest permittivity.
First, we will consider the image mirror obtained at low lecture potential (in the range 200-400 eV). For such low potentials, the point charge approximation is supposed to be valid and the charge will be considered as isotropic. Thus, it corresponds to the previous treated case where the injected charge is a point charge at the interface between a LHOA dielectric and the vacuum. In that case the two mirror diameters $d_{X2}$, $d_{Y2}$, corresponding respectively to the main-major axis and main-minor axis values of the mirror image, should fulfil the condition defined by equation (43) whatever the low lecture potential.
Experimentally, the slope at the origin of the 1/d curves versus $V_l$ corresponding to the two main-axes of the elliptic mirror can be evaluated on figure 8:

$$\frac{1}{d_{\text{major-axis}}} = \frac{1}{d_{X_2}} = 2.82 V_l \quad (47)$$

$$\frac{1}{d_{\text{minor-axis}}} = \frac{1}{d_{Y_2}} = 3.97 V_l \quad (48)$$

The ratio $d_{\text{major-axis}}/d_{\text{minor-axis}}$ is then approximately equal to 1.41 experimentally, that corresponds to the expected analytical ratio $\frac{R_x}{R_y} = \frac{1}{\alpha^2} = \sqrt{\frac{\varepsilon_x}{\varepsilon_y}} = \sqrt{\frac{170}{86}} = 1.406$

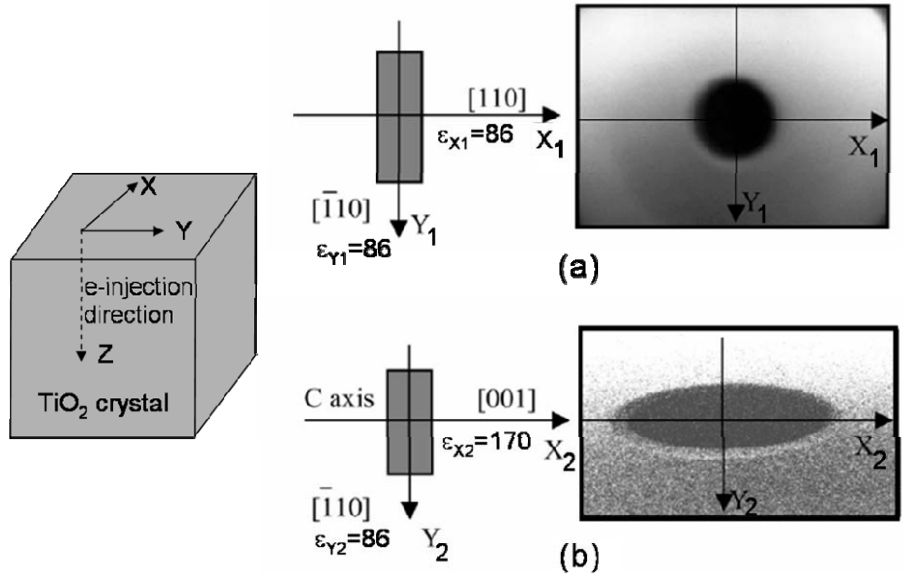

**Figure 7.** Experimental mirror images obtained on $TiO_2$ samples with a 1keV lecture potential in two different e-injection directions: (a) Injection in the [001] direction with the definition of the corresponding coordinate system CS1=$(X_1,Y_1,Z_1)$. Note that the permittivity is isotropic in the $OX_1Y_1$ injection plane (b) Injection in the [110] direction with the definition of the corresponding coordinate system CS2=$(X_2,Y_2,Z_2)$ Note that the permittivity is anisotropic in the $OX_2Y_2$ injection plane.

For higher lecture potential, the point charge approximation is no more verified and a curvature of the curve is observed. This observation is classical and is directly linked to the spreading of the charge distribution for LHI dielectrics [7, 8]. However, for TiO2, one can observe on figure 8 that the variation of the minor-axis diameter is stronger than the ones for the major axis versus the lecture potential value. To interpret these variations of the main-axis value at higher potential, it is possible to combine the two approaches of paragraph 3, anisotropic distribution of the charges, and of paragraph 4, anisotropy of the dielectric. This interpretation necessitates other calculations and will be published later.

.

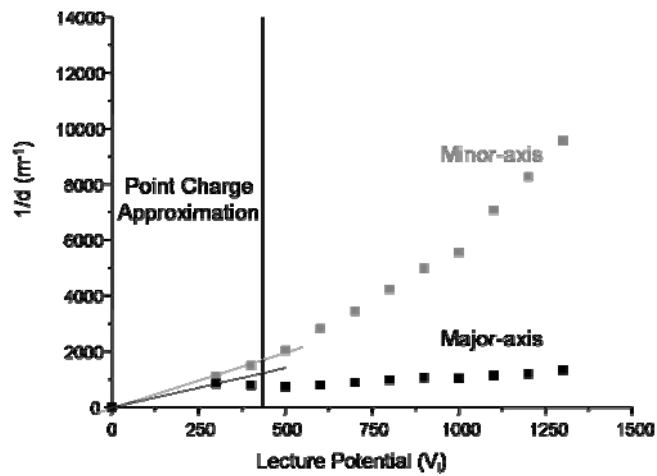

**Figure 8.** Experimental mirror curves $1/d=f(V_l)$ that show the evolutions of the two main-axis diameters of the elliptic mirror obtained on $TiO_2$ in the case of the [110] e-injection versus the lecture potential $V_l$.

## 6. Conclusion

The development of new materials leads to obtain anisotropic structures that have also anisotropic properties (mechanical, optical and dielectrical). This article presents a new analytical method based on linear transformations of the classical electrostatic solutions, in order to calculate the electrostatic field created by electric charges embedded in such anisotropic dielectrics ($\varepsilon_x \neq \varepsilon_y \neq \varepsilon_z$). In this case, the equipotentials are ellipsoids and the electrostatic potential can be evaluated analytically or by finite element simulation.

The anisotropy can also influence the transport of electrical charges and their trapping in the insulators. One relevant way to characterize the transport properties of insulators and their ability to trap electrical charges is the use of the mirror method. However the interpretation of the SEMME experimental results was only known for isotropic materials and for isotropic (or axisymetric) distribution of the stored electrical charges, and was not correct for anisotropic materials.

In this article, we have shown that elliptic mirrors can be obtained analytically either for a homeoidal charge distribution in an isotropic dielectric or for a charge embedded in a LHOA dielectric with two different permittivities in the e-injection plane. It is now possible to quantitatively interpret the experimental mirror results, but also any other effect resulting for a natural or artificial injection of electrical charges in an anisotropic dielectric which results in the creation of a local electrostatic potential.

**Appendix A: Curvature tensor of an ellipsoidal equipotential**

An ellipsoid can be defined by its quadratic form M, where M is a 2x2 matrix, and by the equation

$$^t\vec{x}.M.\vec{x} = 1 \tag{A.1}$$

The tangential plane $\Sigma$ at the point $M_0(\vec{x}_0)$ follows the equation

$$^t\delta\vec{x}.M.\vec{x}_0 = 0 \tag{A.2}$$

where $\delta\vec{x}$ belongs to the tangential plane $\Sigma$. The normal vector $\vec{n}_0$ at the tangential plane $\Sigma$ is given by (cf. figure 2b)

$$\vec{n}_0 = \frac{M.\vec{x}_0}{\|M.\vec{x}_0\|} \tag{A.3}$$

Locally, near $M_0$, a point $M'(\vec{x}')$ with $\vec{x}' = \vec{x}_0 + \delta\vec{x} + z\vec{n}_0$ and $\delta\vec{x} \in \Sigma$ belongs to the ellipsoid if it follows the equation (a). By writing equation (1) for M' up to second order terms in $z^2$, one obtains:

$$z = \frac{-1}{2} \frac{^t\delta\vec{x}.M.\delta\vec{x}}{\|M.\vec{x}_0\|} \tag{A.4}$$

Moreover, the local equation of the surface near the point $M_0$ can also be written versus the local curvature tensor (see figure 2(b))

$$z = \frac{-1}{2} {}^t\delta\vec{x}.C_\Sigma.\delta\vec{x} \tag{A.5}$$

So, we finally obtain

$$C_\Sigma = \frac{M_\Sigma}{\|M.\vec{x}_0\|} = \frac{\begin{pmatrix} M_x & 0 \\ 0 & M_y \end{pmatrix}}{zM_z} = \sqrt{M_z}\begin{pmatrix} M_x & 0 \\ 0 & M_y \end{pmatrix} \tag{A.6}$$

where $M_\Sigma$ and $C_\Sigma$ are respectively the restriction of the quadratic form M to the tangential plane $\Sigma$ and $C_\Sigma$ the curvature tensor at that point.

## Appendix B: Reference electrostatic problem

Consider the case of a point charge q at the interface between two LHOA dielectrics with permittivities $\varepsilon_{1\text{ref}}$ and $\varepsilon_{2\text{ref}}$ defined as follow $\varepsilon_{1\text{ref}} = \begin{pmatrix} \alpha & 0 & 0 \\ 0 & 1/\alpha & 0 \\ 0 & 0 & \beta \end{pmatrix}$ and $\varepsilon_{2\text{ref}} = \begin{pmatrix} 1/\alpha & 0 & 0 \\ 0 & \alpha & 0 \\ 0 & 0 & \beta \end{pmatrix}$

. The equipotentials for such a problem can be approximated in a first step by spheres, and then the potential can be written:

$$V(r) = \frac{Cq}{4\pi\varepsilon_0 (x^2 + y^2 + z^2)^{\frac{1}{2}}} \quad (B.1)$$

For example in the upper half-space, by writing the corresponding field $\vec{D}$ :

$$\vec{D} = \varepsilon_{1\text{ref}} \cdot (-\vec{\nabla} V(r)) = \frac{Cq}{4\pi\varepsilon_0} (x^2 + y^2 + z^2)^{\frac{-3}{2}} \begin{pmatrix} \alpha x \\ y/\alpha \\ \beta z \end{pmatrix} \quad (B.2)$$

This field has to follow the equation $\text{div}(\vec{D}) = 0$ after calculation one obtains: $\text{div}(\vec{D}) = \frac{Cq}{4\pi\varepsilon_0} (x^2 + y^2 + z^2)^{\frac{-5}{2}} (Ix^2 + Jy^2 + Kz^2) = 0$ with $I = -2\alpha + \frac{1}{\alpha} + \beta$, $J = \alpha - \frac{2}{\alpha} + \beta$ and $K = \alpha + \frac{1}{\alpha} - 2\beta$. To have the exact solution we should have I = J = K = 0, that is not possible. However in the case of the backscattering of electrons on an equipotential near the e-beam axis, the important term is the one concerning the space coordinate z that must be minimized. So by choosing $\beta = \frac{1}{2}\left(\alpha + \frac{1}{\alpha}\right)$ the term K is zero, and $I = -J = (\frac{1}{\alpha} - \alpha)$. With this solution there is no increase of the error when z is increasing and the errors in the plane z=0 are self-compensating because of the exchange between $\alpha$ and $1/\alpha$ and they are null on the diagonals x = ±y. The same calculations can be done in the lower half-space. Finally, the corresponding charge q is given by

$$q = \int_{\substack{\text{upper} \\ \text{half-space}}} \vec{D}.\vec{n} + \int_{\substack{\text{lower} \\ \text{half-space}}} \vec{D}.\vec{n} = C\beta q \quad (B.3)$$

which determines the constant C=1/β.
By replacing the constant C in the equation (g), it comes:

$$V(r) = \frac{q}{4\pi\varepsilon_0 r} \frac{1}{\beta} \quad (B.4)$$